\documentclass{aa}  

\usepackage{graphicx}
\usepackage{txfonts}
\usepackage{grffile}
\usepackage{txfonts}

\def\ch{{\it Chandra}}

\def\at{{\it Athena}}

\def\xmm{XMM-{\it Newton}}

\def\Halpha{\ifmmode {\rm H}\alpha \else H$\alpha$\fi}
\def\Hbeta{\ifmmode {\rm H}\beta \else H$\beta$\fi}
\def\Hgamma{\ifmmode {\rm H}\gamma \else H$\gamma$\fi}
\def\Hdelta{\ifmmode {\rm H}\delta \else H$\delta$\fi}
\def\Lya{\ifmmode {\rm Ly}\alpha \else Ly$\alpha$\fi}
\def\Lyb{\ifmmode {\rm Ly}\beta \else Ly$\beta$\fi}
\def\Lyg{\ifmmode {\rm Ly}\beta \else Ly$\gamma$\fi}

\def\ciii{\ifmmode {\rm C}\,{\sc iii} \else C\,{\sc iii}\fi}
\def\civ{\ifmmode {\rm C}\,{\sc iv} \else C\,{\sc iv}\fi}
\def\cv{\ifmmode {\rm C}\,{\sc v} \else C\,{\sc v}\fi}
\def\cvi{\ifmmode {\rm C}\,{\sc vi} \else C\,{\sc vi}\fi}

\def\o5007{[O\,{\sc iii}]\,$\lambda5007$}

\newcommand{\gx}{GX\,5-1}
\newcommand{\gxt}{GX\,340+00}

\begin{document}

   \title{X-ray extinction from interstellar dust}

   \subtitle{Prospects of observing carbon, sulfur and other trace elements}

   \author{E.~Costantini
          \inst{1},
          S.T.~Zeegers\inst{1,2,3},
		  D.~Rogantini\inst{1},
		  C.P.~de Vries\inst{1},
		  A.G.G.M.~Tielens\inst{2}
		  \and
		  L.B.F.M.~Waters\inst{1,4}
          }

   \institute{\inst{1}SRON, Netherlands Institute for Space Research, Sorbonnelaan, 2, 3584, CA, Utrecht, The Netherlands\\
             \email{e.costantini@sron.nl}\\
  			\inst{2} Leiden Observatory, Leiden University, PO Box 9513 2300 RA Leiden, the Netherlands\\
   			\inst{3} Academia Sinica Institute of Astronomy and Astrophysics, 11F of AS/NTU  Astronomy-Mathematics
			Building, No.1, Section 4, Roosevelt Rd, Taipei10617, Taiwan, ROC\\ 
			\inst{4} Anton Pannekoek Institute, University of Amsterdam, Postbus 94249, 1090 GE Amsterdam, The Netherlands\\
              }

   \date{Received/Accepted}

% \abstract{}{}{}{}{} 
% 5 {} token are mandatory
 
  \abstract
  % context heading (optional)
  % aims heading (mandatory)
   {}{We present a study on the prospects of observing carbon, sulfur, and other lower abundance elements (namely Al, Ca, Ti and Ni) present in the interstellar medium using future 
   X-ray instruments. We focus in
   particular on the detection and characterization of interstellar dust along the lines of sight.}
  % methods heading (mandatory)
   {We compare the simulated data with different sets of dust aggregates, either obtained from past literature or measured by us using the SOLEIL-LUCIA synchrotron beamline. Extinction by
   interstellar grains induces modulations of a given photolelectric edge, which can be in principle traced back to the chemistry of the absorbing grains. We simulated data of instruments
   with characteristics of resolution and sensitivity of the current \at, XRISM and Arcus concepts.}
  % results heading (mandatory)
   {In the relatively near future, the depletion and abundances of the elements under study will be determined with confidence. In the case of carbon and sulfur, the characterization of
   the chemistry of the absorbing dust will be also determined, depending on the dominant compound. 
   For aluminum and calcium, despite the large depletion in the interstellar medium and the prominent
   dust absorption, in many cases the edge feature may not be changing
   significantly with the change of chemistry in the Al- or Ca- bearing compounds. The exinction signature of large grains may be detected and modeled, allowing a
   test on different grain size distributions for these elements. The low cosmic abundance of Ti and Ni will not allow us a detailed study of the edge features.}{}

\keywords{ISM: dust, extinction --  X-rays: ISM -- X-rays: individuals: GX\,5-1, GX\,340+00, GX\,3+1 -- techniques:
spectroscopic}
\authorrunning{E.~Costantini et al.}
\titlerunning{X-ray extinction from interstellar dust}

   \maketitle
%
%-------------------------------------------------------------------

\section{Introduction}

Absorption and scattering in the X-ray band has proved a useful diagnostic of the interstellar dust (ID) properties.
By virtue of the broad band coverage, the X-ray band displays many photoelectric absorption edges, caused by the mixture of
gas and dust intervening along the line of sight towards bright background sources \citep{draine03,hoffman_draine16}. Absorption by interstellar grains is detected as a result of 
the interaction between the incoming X-ray photon and the electrons inside the grain's atoms. 
The multiple-generated photoelectron-waves interfere with each other 
both constructively and destructively. This interference pattern depends on the complexity 
of the chemical compound and the distance of the electrons from the nucleus. 
Each pattern is a fingerprint of a given material \citep{r_a00}.  
The extinction cross section, the sum of the absorption and scattering cross section 
\citep[e.g.][]{draine03,corrales16} in the X-ray band, provides, in principle, not only direct estimate on the chemistry 
of the interstellar medium (ISM), but also information on the size distribution, crystallinity and porosity 
of the intervening grains \citep[][]{hoffman_draine16, zeegers17, rogantini18}.

Early studies already pointed out that absorption by the ISM contributed to the shape of X-ray spectra
\citep{sc86,paerels01,juett04}. However, in recent years, the deep 
features of the Fe~L and O~K and Si~K edges have been recognized to be largely caused by dust absorption
\citep[e.g.][]{lee02, ueda05, schulz16} and have been studied using the grating spectrometers 
on board the X-ray Observatories \ch\ and \xmm. These studies made use of 
absorption profiles either taken from the literature \citep[][]{costantini12,pinto10,pinto13,valencic13} 
or obtained with dedicated synchrotron measurements \citep{lee09, zeegers17}.   
\begin{center}
\begin{figure*}\label{f:abu}
\includegraphics[width=12cm, height=18cm,angle=90]{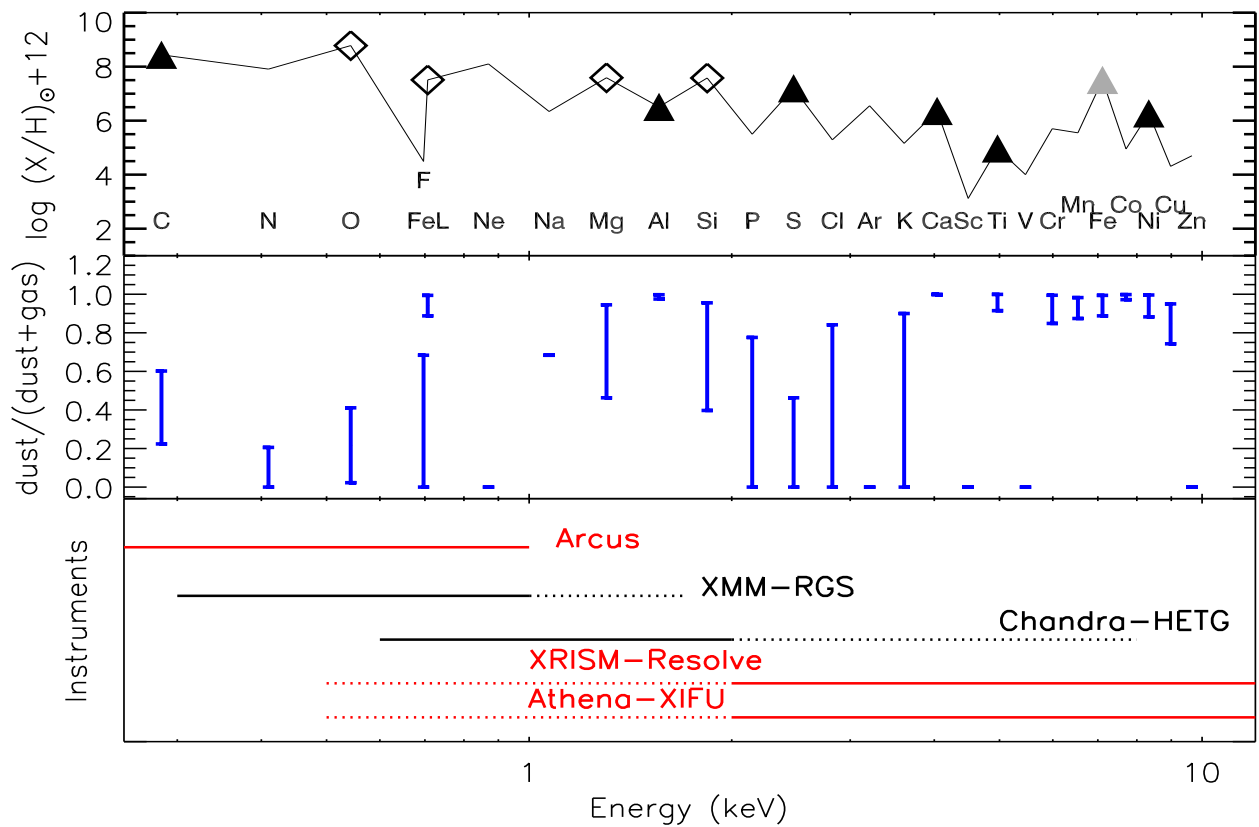}
\caption{Upper panel: abundance pattern as a function of energy for the absorbing elements in the X-ray band. The K-edge energy is indicated, except from Fe, for which both the K- and
the L-edges, at 7.1 and $\sim$0.7\,keV, respectively, can be studied. Abundances
follow \citet{lodders10} and they are expressed in terms of log (X/H)+12. In this frame, the abundance of hydrogen is 12.
The open diamond mark the elements that are accessible by current instruments. The triangles are the relevant elements
that will be accessible by future instruments to study dust. The black triangles are the subject of this work.  
Middle panel: range of depletions as reported by \citet{jenkins09} for all elements, except: C \citep{jenkins09,whittet03}, F \citep{snow07}, Na
\citep{turner91}, S \citep{gry17}, K \citep{snow75}, Ca \citep{crinklaw94}, Co \citep{federman93}, Al \citep{jenkins96}, Ar \citep{sofia_jenkins98}. 
Lower panel: energy range covered by present and future (red) mission. The solid line highlights the energy range
where the instruments capabilities are optimal for observing absorption by dust.}
\end{figure*}
\end{center}

%The carbon K~edge has been studied in
%\citep{ss10}, using the \ch-LETGS. The view on carbon is less clear due to relatively high sky background at 0.28\,keV
%and other effects specific to that instrument (namely the 
%contamination by the carbon in the instrument itself and the superposition of grating orders at that energy).       

Outside the energy band where the sensitivity and resolution of the current instruments is maximized, it is at this
moment challenging to study interstellar dust. An example is given by the tentative study of the C~K edge \citep{ss10},
which was severely hampered by various instrumental effects, although the carbon edge would formally be included 
in the energy range of \ch-LETGS.

In this paper we investigate the prospective of observing and modelling the 
elements of the ID which have not been yet studied, which happen to fall in the 1.5--8.3\,keV band (Al, S, Ca, Ti, Ni) 
and at E$<$0.5\,keV (carbon). For discussion on how the features that can be currently studied (O\,K,
Fe\,L Mg\,K, and Si\,K edges) will be viewed by future instruments, we refer to e.g. \citet{decourchelle13,smith16}.

In Fig.~\ref{f:abu} we show the abundance pattern of the photoelectric 
absorption edges of the elements (with atomic number A=6-30) as a function of the X-ray energy. The empty diamonds mark
the edges that have been already studied by current instruments unveiling the dust features: the O K and Fe L edges
at 0.534 and 0.7\,keV respectively \citep{lee09,costantini12,pinto10,pinto13,valencic13}; the Mg and Si-K edges at
1.3 and 1.84\,keV respectively \citep[][Rogantini et al. in prep.]{zeegers17, zeegers19}. The black triangles mark the edges
presented in this work. We present the Fe~K edge, marked with a light gray triangle in the figure, in a separate paper
\citep{rogantini18}, but see also \citet{lee05}.
The middle panel shows the range of depletion, defined as the amount of dust over the total amount of matter in the ISM
(dust and gas) that is expected for a given element. The wide range of depletions for some elements 
is due to the different density environments where those are observed \citep[e.g.][]{jenkins09}.       

In the lower panel of Fig.~\ref{f:abu} we show the energy range of present and future missions. The solid line highlights the region
where the instrument capabilities (resolution and effective area) are optimal to observe the dust absorption features. 
The \ch\ and \xmm\ observatory \citep[both launched in 1999, ][]{w99,j01} are still in operation. 
The grating spectrometer Arcus will cover the soft X-ray range (Table~\ref{t:reso}). It is a NASA mission 
currently in the study phase \citep{smith16}. The calorimeter on board the X-ray
Imaging Spectroscopy Mission 
(XRISM, to be launched around 2021) is  planned to have the same characteristics of the one on board of the lost {\it Hitomi} satellite \citep{mitsuda14}. 
Finally, we display the energy coverage of the \at\ calorimeter XIFU \citep{barret16}, to be launched in 2030. Both XRISM and \at's calorimeter will be optimal
to observe the higher energy dust features (Table~\ref{t:reso}).
\begin{table}
\caption{\label{t:reso} Parameters of the instruments used in the simulations at the energy of the elements studied here.}
\begin{center}
\begin{tabular}{llll}
\hline
\hline
Element & R & A$^{Eff}_{E}$ & Inst.\\
& E/$\Delta$E & cm$^{2}$ & \\
\hline
C & 2540 & 369 & Arcus\\
Al & 621 & 11022 & XIFU\\ 
S & 986(352) & 5949(209) & XIFU(Resolve)\\
Ca & 1611(575) & 3734(271) & XIFU(Resolve)\\
Ti & 1984 & 2742& XIFU\\
Ni & 3331 & 704& XIFU\\
\hline
\end{tabular}
\end{center}
Notes: the energy resolution is 2.5 and 5 eV for XIFU  and Resolve, respectively. Arcus resolution is defined by its resolving power
R$\sim$3000 over the 0.5--1 keV band. 
\end{table}

\subsection{The elements in this study}\label{par:elements}

One of the major players in the ID, carbon, constitutes around 20\% of the total depleted mass in the
Galaxy \citep{whittet03}. Its depletion covers a relatively narrow range of values, (Figure~\ref{f:abu}) showing that it is 
not a strong function of environmental density. It has been hypothesized that the majority of carbon should be locked in
graphite grains, as a likely explanation for the 2175\,\AA\ emission feature 
\citep[][and references therein]{draine89, draine03}. Graphite has been commonly adopted in ID models \citep[e.g.][]{mrn77}. However,
observational evidences pointed out that graphite could not explain the variability of the 2175\,\AA\ feature \citep[][]{fitzpatrick_massa07}. 
Furthermore, in analogy with the silicates, which are found to be amorphous, also graphite 
was deemed unlikely to survive in large quantities in the harsh
environment of the ISM. Graphite should therefore face a natural process of amorphisation \citep{compiegne11}.    

The idea of carbon as a single and separate phase from the silicate population does not agree with a scenario of a constantly evolving and mixing medium \citep{jones17}. Hydrogenated
amorphous carbon (HAC) may indeed coat the silicate grains, forming a single population \citep[e.g.][]{duley89}, 
with different characteristics with respect to the environment where they reside and depending on the particle size  
\citep[e.g.][and references therein]{jones17}. Polarization studies however did not confirm so far this scenario. The carbon feature at 3.4$\,\mu$m shows a
negligible degree of polarization with respect to the silicate feature at 9.7$\mu$m, pointing to two distinct grain populations \citep{whittet11}. 
%Large carbon grains may be more
%hydrogenated as less affected by heating. On the contrary, in smaller grains, hydrogen may be eroded away in high-temperature environments
%\citep[e.g.][]{jones09}. 

Finally, under special condition of high pressure, for instance in a shocked environment, graphite and amorphous carbon can turn into nano-diamonds, which can constitute as much as 5\% of
the amount of C in the ISM \citep{tielens87}, possibly with H and N inclusion \citep{kerck02, goranka18}. Diamonds of possible ISM origin have been found in meteorites \citep{lewis87}.     
An important carbon carrier are Polycyclic Aromatic Hydrocarbons (PAH), large molecules (\AA ngstrom-sized) formed by 
carbon and hydrogen in a honeycomb structure. They constitute up to about 10\% of the carbon abundance \citep[e.g.][]{tielens13}. PAHs are quite
sensitive to ionizing radiation from far-ultraviolet to X-rays and they are easily destroyed near star formation sites 
 at AU distance scale \citep[e.g.][]{siebenmorgen10}, up to kpc scale, for active galaxies \citep[][]{voit92}.  

Apart from C, other constituents can be studied in detail by future
generation telescopes. 

Sulfur in dust phase seems to be absent from the diffuse ISM \citep{ss96}. However, a relative 
fast transition to a depletion approaching -1\,dex is reported in dense media, such as molecular clouds \citep{joseph86}. In molecular clouds, sulfur can be included 
in aggregates such as H2S, SO2, OCS, SO,
H2CS, NS SiS, CS, HNCS CH3SH \citep[and references therein]{duley80} as well as other carbon-hydrogen bearing sulfates \citep[e.g.][]{goranka15}. 
Molecular reactions may also lead to sulfur aggregation into polymeric forms, like S$_8$
\citep[e.g.][]{jimenez11}. However, even integrating the contribution of all S-bearing molecules, the absolute abundance of sulfur 
in molecular clouds compared to the diffuse ISM one, with a ratio of $\sim 10^{-8}/10^{-5}$, is inexplicably low \citep{wakelam08}. Inclusion into simple atomic
sulfur or sulfur ices have been proposed to solve the missing-sulfur problem in molecular clouds \citep[e.g.][]{vidal17}.\\
Sulfur in dust has been also detected near C-rich AGB stars, planetary nebulae \citep{hony02} and protoplanetary
disks \citep{keller02}, predominantly in form of troilite (FeS). Finally, sulfur is abundant in solid form in planetary systems bodies, such as 
interplanetary dust particles, meteorites and comets \citep[e.g.][ and references therein]{wooden08}.\\
The presence of sulfur in dust form in the ISM has been suggested in association with GEMS \citep[Glasses with Embedded Metal and Sulfides,][]{bradley94}, where the FeS
particles would be more concentrated on the surface of the glassy silicate. However, the majority of GEMS may well be of nebular origin, rather than the ISM
\citep{keller_messenger08}.
Sulfur in FeS, consistent to be of ISM origin, has been recorded in the data from the Stardust mission \citep{westphal14}. This evidence revitalizes 
the idea of the presence of sulfur in dust-form also in less dense environments of the ISM. The presence of strong
UV radiation and cosmic rays has been thought to be the cause for the extreme sputtering of the highly volatile 
S, for example from GEMS surfaces. Recent experiments however put to the test this hypothesis \citep{keller10}, showing that UV
bombardment has in fact little influence on sulfur stuck on a grain surface. 
   
\begin{table*}
\caption{\label{t:sample}Samples of interstellar dust analogues used in this work} 
\begin{tabular}{llcc}
\hline
\hline
Specie & Name & atom& Ref \\
C & {\small graphite} & C & {\small \citet{gago98}} \\
AC & {\small amorphous carbon} & C& {\small\citet{gago98}}\\
HAC & {\small hydrogenated amorphous carbon} &C& {\small\citet{buijnsters09}}\\
C & {\small diamond} & C &{\small\citet{gago98}}\\

MgAl$_2$O$_4$ & {\small spinel} & Al & {\small this work} \\
Al$_2$O$_3$ & {\small aluminum oxide} & Al& {\small this work}\\

%MnS & {\small alabandite} & S & {\small\citet{bonnin02}} & 2476 & 2472\\
FeS$_2$ & {\small pyrite}&S &  {\small\citet{bonnin02}} \\
FeS &{\small  troilite}&S & $^1$ \\
Fe$_{0.875}$S & {\small pyrrohtite}&S & $^1$\\

CaMgSi$_2$O$_6$ & {\small diopside crystal} & Ca & {\small\citet{neuville07}} \\ 
CaMgSi$_2$O$_6$ & {\small diopside glass} & Ca &{\small\citet{neuville07}}\\ 
Ca$_3$Al$_2$O$_6$ & {\small tricalcium aluminate} & Ca & {\small\citet{neuville07}}\\
CaAl$_2$O$_4$ &{\small calcium aluminate}&Ca & {\small\citet{neuville07}}\\

TiO$_2$ & {\small titanium dioxide} & Ti & {\small\citet{shin13}} \\  

Ni &{\small  metallic nickel} & Ni &  {\small\citet{vanloon15}} \\
\hline

\end{tabular}
{\small 
\newline
$^1$ http://www.esrf.eu/home/UsersAndScience/Experiments/XNP/ID21/php/Database-SCompounds.html}\\
\end{table*}

Aluminum, calcium and titanium are extremely depleted in the ISM (Fig.~\ref{f:abu}). 
Ca and Ti show also a very similar depletion pattern as a function of gas density \citep{crinklaw94}. 
These two elements are found in gas only in tenuous environments, associated with warm inter cloud media in both the halo \citep{edgar_savage89} 
and the disk of the Galaxy \citep{crinklaw94}. The depletion of Ti is severe, regardless of the environment, ranging between -1\,dex and -3.1\,dex \citep{welty_crowther10}. 
The ratio of column density between \ion{Ti}{ii} and \ion{Ca}{ii}, representative of the element neutral gas phase, 
is in general constant in the Galaxy \citep[$\sim$0.4,][]{hunter06}.
Al, Ca and Ti have a similar condensation temperature \citep[1400--1600\,K,][]{field74} and it has been hypothesized that, being the first to form in e.g. a stellar envelope or a
supernova environment, they would form the core of complex dust grains with silicate and possibly ice mantles \citep[e.g.][]{clayton78}. 
This would provide a natural shield for these Al, Ca, Ti-bearing compounds, preventing their destruction and ensuring a high depletion 
in the vast majority of the environments. 
Under the condition of thermodynamic equilibrium, aluminum first condenses in Al$_2$O$_3$. From there it may 
evolve into spinel (MgAl$_2$O$_4$) and eventually into a Ca and Al-bearing silicate. The latter are stable compounds, thanks to very high binding energies \citep[][]{trivedi81}. 
Calcium is mostly locked in dust in silicates \citep[e.g. CaMgSi$_2$O$_6$,][]{field74, trivedi81}. Calcium carbonates, 
possibly formed in AGB stars envelopes \citep[e.g.][]{kemper02}, are believed to be unstable and therefore Ca inclusion in silicates, which form 
already at high temperatures, are favored \citep{ferrarotti05}. \\
Titanium is produced by AGB stars mostly in the form of TiO$_{2}$, which constitute a seed nucleus later included in the larger/coated grains  \citep[e.g.][]{ferrarotti_gail06}. 

Nickel depletion is found to correlate with the one of Fe for a variety of environments, 
from planetary nebulae \citep{di16} to diffuse interstellar clouds \citep{jenkins09}. These two
elements display a similar condensation temperature \citep[1336 and 1354\,K for Fe and Ni, respectively,][]{wasson85},
which already point to a simultaneous inclusion into dust grains. However, it has been observed that in dense environments
Ni is more depleted than Fe \citep[e.g.][]{ss96,di16}. 

\section{Extinction profiles}
In this paper, we make use of literature values to infer the absorption absolute cross sections for all elements, except for Al (Sect.~\ref{par:lab}). 
Measurements of X-ray edges profiles are mostly carried out for industry and are rarely of interest for astronomical applications. 
For this reason, the sample selection is bound to be incomplete. The compounds used are 
listed in Table~\ref{t:sample}. We follow closely the method presented in \citet{zeegers17} and \citet{rogantini18} 
to obtain the extinction profiles to be confronted to the astronomical
 data. The laboratory data are transformed to transmission spectra and matched (via $\chi^2$ fitting) 
 to tabulated transmission data of the same compound, where we assume an optically thin sample, 
 as to mimic the conditions in the ISM. In doing this we only fit the pre- and post-edge of the tabulated data, leaving
 the edge energy as measured in the laboratory. The transmission tables are provided by the Center for 
 X-ray Optics at Lawrence Berkeley National laboratory based on tabulated data by \citet{henke93}. 
 From the transmission spectra, the attenuation coefficient can be obtained and consequently 
 we can obtain the imaginary part of the refractive index from this coefficient. The real part of 
 the refractive index $m$, is calculated via the Kramers-Kroning relations~\citep{Bohren10}. The 
 knowledge of $m$ is needed in order to calculate the extinction cross section to involve both 
 the effect of absorption and scattering. The extinction cross sections are calculated using Mie theory \citep{mie08,wiscombe80} for C, Al and S. We used instead the anomalous diffraction theory 
 (ADT, Van der Hulst 1957) for Ca, Ti and Ni. The ADT theory can be used when the ratio 
 $x = 2\pi a/\lambda\gg 1$ where $a$ is the grain size and $\lambda$ is the wavelength of the incident radiation. 
 In order to obtain the extinction cross section for a range of dust grain radii, we assume the MRN size distribution 
 (Mathis et al. 1977 and Sect.~\ref{par:simu}). Once the absolute cross section as a function of energy has been obtained, we 
 implemented the extinction profiles in the already existing AMOL model in the fitting code SPEX 
 \citep[ver. 3.03,][]{kaastra96}. The AMOL model is an absorption model to be applied to the emitting continuum
 model of a source. It allows to fit the X-ray edges 
 for the column densities of a set of four dust compounds at a time \citep[see also][]{zeegers17}. 
 In real astronomical observations, the dust extinction feature will always coexist with the gas feature of the
corresponding element. The absolute energy of the edges of the gas phase may be reported in the literature at different
energies, with discrepancies sometimes of few eV. In SPEX, the gas edge energies,
 for the elements in this work, are implemented following \citet{verner96}. In this paper, 
 we apply a shift to the laboratory data in order to consistently compare them with the gas edge 
 features as seen by SPEX. Discrepancies among different measurements and theoretical calculations may be found in the literature. 
 High resolution X-ray spectroscopy will help determining the absolute energy scale of the edge \citep[e.g.][]{gorczyca}.  

\subsection{Laboratory data for Al}\label{par:lab}

For Al, we made use of the laboratory data that we collected at the LUCIA beamline at the Soleil synchrotron facility which offers
an energy resolution of $\sim$0.25 eV. Both the samples, Al$_2$O$_3$ and MgAl$_2$O$_4$, were commercially available from the Alfa-Aesar and Aldrich company, respectively.  
The samples, in powder form, were pressed on thin indium foil, placed on a copper support which was placed in a vacuum environment. The sample 
was then irradiated by an X-ray beam of which the energy is tuneable. The X-ray fine structures were measured through fluorescence. 
At these soft X-ray energies, this method is more practical than the more intuitive method of measuring the transmission 
through the sample, because for transmission measurements the samples have to be too thin to be easily handled. 
The fluorescent method to obtain the XAFS does require a correction for possible saturation. This correction was performed 
with the program FLUO, which is part of the UWXAFS software \citep{stern95}. 
A full description of the procedure for the analysis of the data can be found in \citet{zeegers17}.

%{\small
%}
\section{Simulations}\label{par:simu}

We present the prospects of detecting absorption K edges relevant for dust studies using future missions (Fig.~\ref{f:abu}). 
The only instrument proposed for studying the 
soft X-ray energy band is, at this moment, the Arcus grating spectrometer\footnote{http://www.arcusxray.org/} \citep{smith16}. 
For the energy above $\sim$2\,keV, two microcalorimeters will provide an unprecedented resolution: {\it Resolve} and XIFU, on board of 
XRISM\footnote{https://heasarc.gsfc.nasa.gov/docs/xrism/} and 
\at\footnote{http://www.the-athena-x-ray-observatory.eu/} \citep{nandra13}, respectively. 
The effective area and the resolving power of these three instruments at the energy of the features studied here are reported in
Table~\ref{t:reso}. With the chosen exposure time we would obtain an associated error on the dust or gas column density of around 1\% for C (Arcus) Al, S
and Ca (XIFU).  

The simulations were carried out having in mind realistic scenarios in our Galaxy, 
in order to prove the effective prospect of future instruments to measure physical parameters.
We first simulated the data, considering the different instruments responses and including noise, 
assuming first that the photoelectric edge is only due to gas absorption.   
These simple simulations are confronted with models, folded with the appropriate response, which include 
an amount of gas set by the typical depletion found in the literature for a given element plus the contribution of
a dust compound. We adopted the MRN dust size distribution in all cases \citep{mrn77}. The MRN model offers a simple
parameterization of the dust size distribution: $n(a)\propto a^{-3.5}$. It has been found to approximate at first order the conditions in
our Galaxy for grain size between 0.025 and 0.25$\mu$m. 
However, the exceptional depletion of Al and Ca, 
joint to favorable observing conditions allowed us to test also the detectability of a distribution where the mass distribution is skewed towards larger grains
\citep{draine_f}. This distribution has a size range of $a=0.02-1$\,$\mu$m, with an average grain size of 0.6\,$\mu$m. 
The observing conditions are favorable for these two elements first because the brightness of the sources with a favorable $N_{\rm H}$ is
often high (e.g. low mass X-ray binaries, LMXB, near the galactic center) and, second, the edges fall in a large effective-area region of the instrument.\\
Unless otherwise stated, we chosed the maximum depletion allowed by previous studies (Tab.~\ref{f:abu}). 
This is  a reasonable
assumption in most cases. Indeed, for the photoelectric edge to be detected, a substantial column density is required. 
This also indicate a dense environment, where the depletion is large. Here 
below we describe the conditions under which the simulations were performed for each element.

{\bf Carbon:} the depletion of carbon has been optimistically assumed to be 0.6, but still implies a substantial role in gas absorption in the carbon edge. For this
reason, although the edge feature itself can also be detected at relatively small column density (e.g. $N_{\rm H}\textgreater 10^{20}$\,cm$^{-2}$),
a much larger amount of matter is necessary to make the dust features evident. Here we simulated a column density of $N_{\rm H}=1.6\times
10^{21}$\,cm$^{-2}$ for a flux in the soft energy band of $\sim3\times10^{-9}$\,erg\,cm$^{-2}$\,s$^{-1}$ (Fig.~\ref{f:carbon}). 
Note that for this column density, the \ion{C}{i} absorption lines from gas are already saturated (Fig.~\ref{f:carbon}), therefore they cannot be used straightforwardly to 
measure depletion.  Although this value for a column
density is not uncommon for LMXB, the source needs to be in an hypothetical high state to be well detected by an Arcus-like instrument as 
the effective area of such an instrument would fall rapidly at the carbon edge.

{\bf Aluminum: }the cosmic abundance of Al is significantly less than the main ID components (Fig.~\ref{f:abu}). However, Al in the ISM 
is almost completely depleted onto dust. We simulate here the contribution of Al$_2$O$_3$ and MgAl$_2$O$_4$ and compare them with a theoretical pure gas
absorption (Fig.~\ref{f:alluminum}). At ISM temperature Al, if it was totally in gas form, 
would be distributed between \ion{Al}{i} and \ion{Al}{ii}. 
For this simulation we selected the bright LMXB 
($F_{2-10\,{\rm keV}}\sim3\times10^{-9}$\,erg\,cm$^{-2}$\,s$^{-1}$) \object{GX~3+1} whose spectral parameters where obtained from \ch-HETGS data
(obsid 16492).

\begin{figure}
\begin{center}
\resizebox{\hsize}{!}{\includegraphics[angle=90]{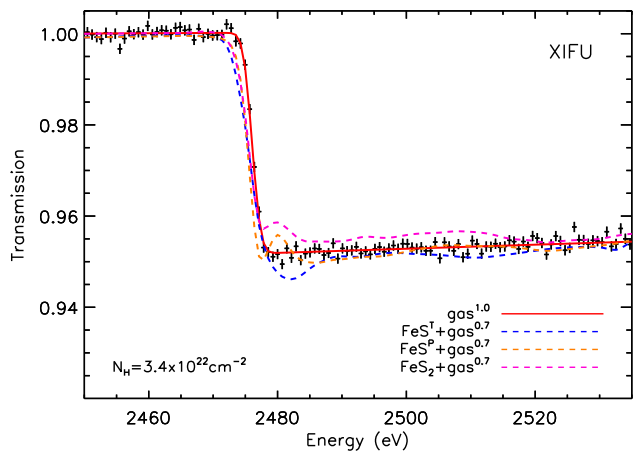}}
\resizebox{\hsize}{!}{\includegraphics[angle=90]{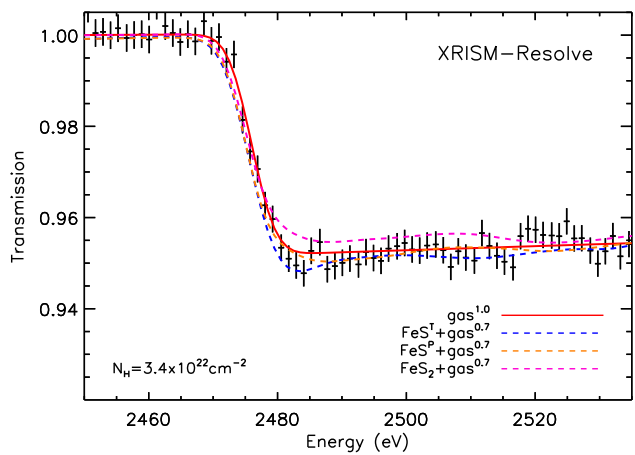}}
\end{center}
\caption{\label{f:sulfur} Simulation of the sulfur K\,edge, using the XIFU calorimeter with an exposure time of 
200\,ks (top) and XRISM-Resolve with a 400\,ks exposure time (bottom), of a the bright XRB GX\,5--1 
($F_{2-10\,{\rm keV}}\sim2.5\times10^{-8}$\,erg\,cm$^{-2}$\,s$^{-1}$). The simulation
considers different sulfur species, with a dust depletion of 30\%.} 
\end{figure}

{\bf Sulfur:} for the sulfur simulation, we selected \object{GX~5-1}, which is among the brightest LMXB in the Galaxy, 
with a $F_{2-10\,{\rm keV}}\sim2.5\times10^{-8}$\,erg\,cm$^{-2}$\,s$^{-1}$. The hydrogen column density is about
$3.4\times10^{22}$\,cm$^{-2}$ \citep[][]{zeegers17}. The depletion of sulfur is unknown in the diffuse ISM, but it 
has been estimated that could be up to $\sim$46\%
\citep[Fig.~\ref{f:abu} and][]{gry17}. Here we simulate a more conservative 30\% depletion. 
Given the relatively low depletion of S, the XAFS features (Fig.~\ref{f:a1}1) would be less evident in the data
(Fig.~\ref{f:sulfur}).         

{\bf Calcium:} The X-ray spectrum will be sensitive to calcium extinction only if the intervening column density is sufficiently high. This is due to 
the relatively high energy position of the photoelectric edge. In Fig.~\ref{f:calcium} 
we simulated \object{GX~340+00} ($F_{2-10\,{\rm keV}}\sim2.5\times10^{-8}$\,erg\,cm$^{-2}$\,s$^{-1}$), with a column density of
about $N_{\rm H}\sim6.9\times10^{22}$\,cm$^{-2}$, obtained from HETG-\ch\ data fitting (obsid 6632).

{\bf Titanium and nickel: }for both these elements, we simulated an hypothetical source, for example near the GC, where also the occurrence of high column density molecular
clouds is more frequent, that in outburst reaches a flux as high as \gxt\ (Fig.~\ref{f:titanium}, \ref{f:nickel}). 
The column density must be sufficient to produce an
edge-like modulation in the spectrum ($N_{\rm H}\sim 1.3\times10^{23}$\,cm$^{-2}$).

\section{Discussion}

\subsection{Carbon}
In the simulation we included gas and the carbon forms
that are believed to be most abundant (namely graphite, amorphous carbon and HAC). 
While the difference between graphite and amorphous carbon is
subtle, HAC does have more distinctive features that may be more easily detected. 
The hydrogenation of carbon may point to either an environment protected from strong radiation or the presence of large
grains, which are more resilient to radiation (Sect.~\ref{par:elements}). 

\begin{figure}
\begin{center}
\resizebox{\hsize}{!}{\includegraphics[angle=90]{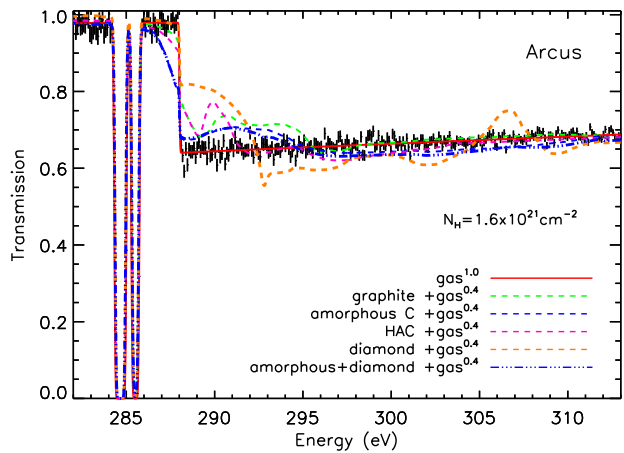}}
\end{center}
\caption{\label{f:carbon} A 500\,ks simulation of the carbon K\,edge, using the Arcus grating, of an XRB in high state ($F_{0.5-2\,{\rm keV}}\sim3\times10^{-9}$\,erg\,cm$^{-2}$\,s$^{-1}$). The simulation
considers different carbon species, with a dust depletion of 60\%. The two absorption lines belong to the atomic phase of C, namely \ion{C}{i}. } 
\end{figure}

For illustrative purposes, we also include diamonds (orange dashed line in Fig.~\ref{f:carbon}) 
in order to show the
departure of this form of carbon from the shape of e.g. graphite. However, 
in practice, diamonds are believed to constitute no more than 5\% of the 
carbon \citep{tielens01}. 
Its realistic inclusion would be non detectable (dashed-dotted blue line in figure). The same negligible effect may be produced by
PAH, which we did not include in our simulation. The total amount along a line of sight is relatively low
(Sect.~\ref{par:elements}) and the spectral features of PAH would be mixed with a more dominant amorphous
carbon (or graphite) contribution. The sparse historical studies of PAH absorption profiles  
for the X-ray region have been recently revived \citep{reitsma,reitsma2}. This will help in defining a shape for the
summed contribution of the numerous different PAH in the ISM.\\ The carbon edge is not very sensitive to the size
distribution of the grains \citep{draine03}, therefore the MRN distribution may be adequate to describe the edge. 
%Moreover, given the relatively low depletion, any scattering peak would be easily diluted in the spectrum.  

\subsection{Aluminum and Calcium}
The XIFU simulation shows that dust will be easily detectable, even if the edge itself produced a jump in the spectrum of only few \%. 
However, from the Al edge alone, it would be difficult to distinguish among different compounds. 

\begin{figure}
\begin{center}
\resizebox{\hsize}{!}{\includegraphics[angle=90]{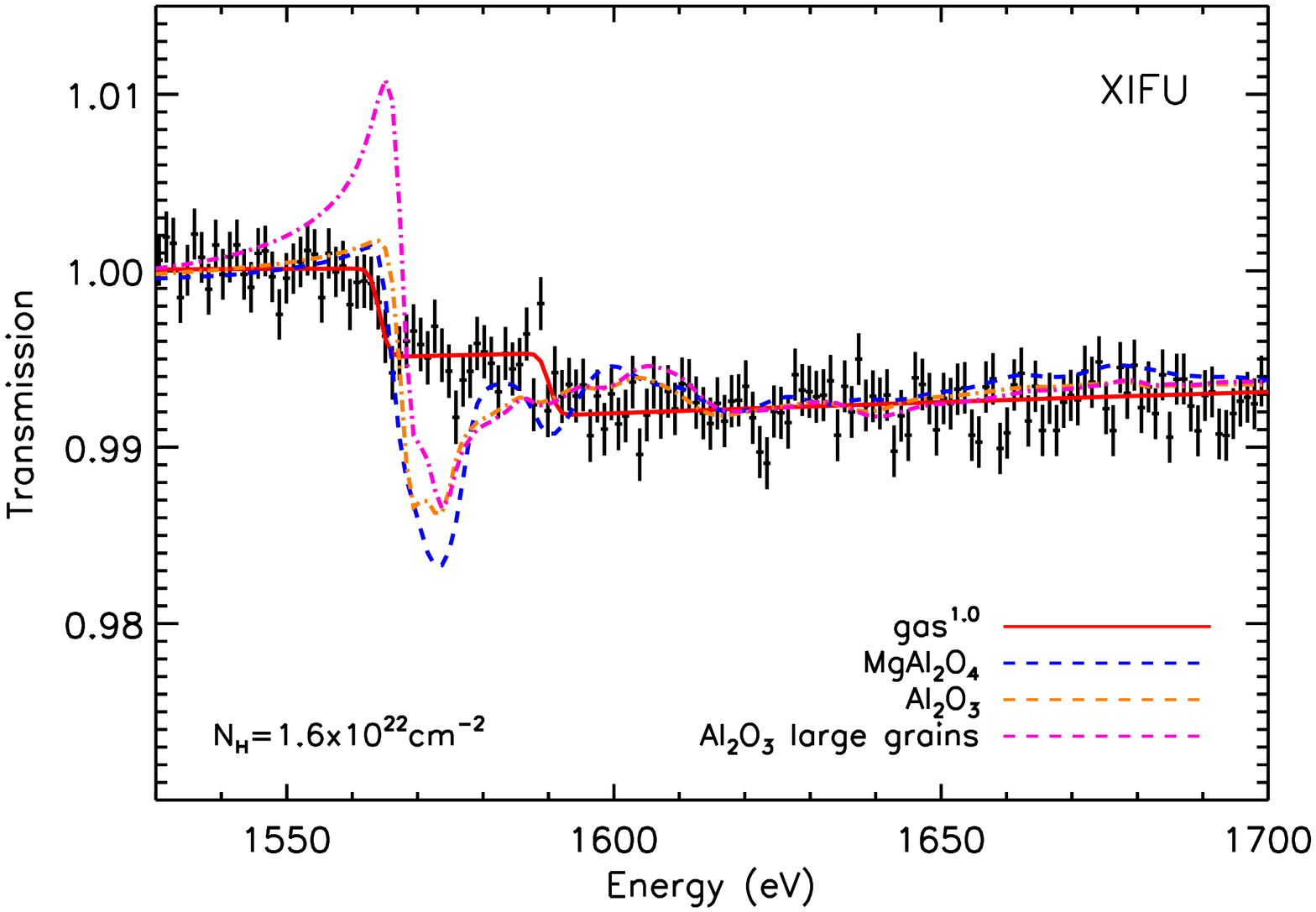}}
\end{center}
\caption{\label{f:alluminum} A 300\,ks simulation of the aluminum K\,edge, using the XIFU calorimeter, of the bright XRB GX\,3+1 
($F_{2-10\,{\rm keV}}\sim3\times10^{-9}$\,erg\,cm$^{-2}$\,s$^{-1}$). The dust depletion is 100\%. The data have been binned for clarity.} 
\end{figure}

\begin{figure}
\begin{center}
\resizebox{\hsize}{!}{\includegraphics[angle=90]{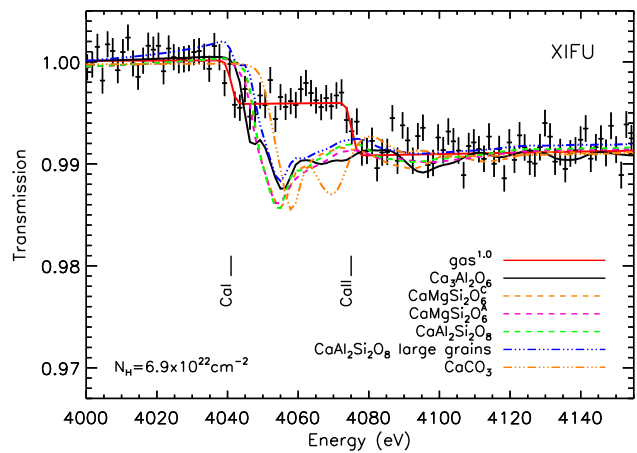}}
\resizebox{\hsize}{!}{\includegraphics[angle=90]{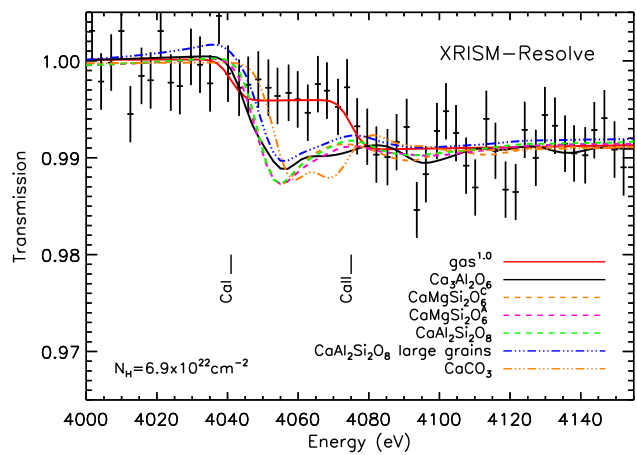}}
\end{center}
\caption{\label{f:calcium} Simulation of the calcium K\,edge, using the XIFU calorimeter with 
exposure time 400\,ks, (top) and the XRISM-Resolve with a 500\,ks exposure time (bottom). We used the bright XRB GX\,340+00 
($F_{2-10\,{\rm keV}}\sim1.3\times10^{-8}$\,erg\,cm$^{-2}$\,s$^{-1}$). The simulation
considers different calcium species, with a dust depletion of 100\%. The data have been binned for clarity.} 
\end{figure}

We also note that contrary to other extinction profiles, the scattering peak, which appears as an
emission-like feature before the edge-jump, is noticeable in Al. This peak is sensitive to the dust size distribution \citep{zeegers17} and
can be used, in principle, to estimate e.g. the mean grain size along the line of sight. As described above, we also tested the effect of the dust size
distribution of \citet{draine_f} for Al$_2$O$_3$. The edge energy of Al lies in a zone sensitive to scattering \citep{draine03}, therefore in Fig.~\ref{f:alluminum} the
large particles contribution is evident. As shown in \citet[][]{rogantini18}, the role of a substantial scattering contribution to the extinction not only
forces the edge energy to shift, but may also modify the appearance of the edge absorption features. Grains containing seeds of Al and Ca, which are shielded from erosion in
the ISM, are believed to be of large size, due to the several layers of coatings surrounding those seeds elements \citep[e.g.][and Sect.~\ref{par:elements}]{clayton78}.
With future instruments we will therefore be able to test also the presence of larger particles for less abundant, but important, constituents of the ISM.
The study of the Al edge will be however challenging, 
as Al is always a major component of X-ray space instruments (often in the form of foils). The extinction feature from Al in the ISM
will be always blended with a relatively deep instrumental Al feature. This would need a careful calibration, adding uncertainty to the modeling.\\  
 Calcium is totally
depleted in the ISM, therefore the main dust features will be detected (Fig.~\ref{f:calcium}). However, calcium is mostly contained in silicates
and aluminates, where oxygen is the main constituent. XAFS models shows that the first and main absorption feature is due to the
nearest neighboring atom that the photoelectron wave will encounter \citep[][]{lee05}. In the case simulated here, the absorption profile
is dominated by oxygen (as is the case, to a lesser extend, in the Al edge), 
and only at higher energies are the secondary absorption features to be seen. For this reason the Ca inclusion in 
a specific silicate may be hard to disentangle through the observed spectrum. However, calcite (CaCO$_3$, dashed 
orange line), due to its different internal structure, will show a distinctive pattern, which may be in 
principle disentangled. This will help in determining whether this elusive compound \citep[e.g.][and Sect.~\ref{par:elements}]{kemper02} may be present in the ISM. 
We tested the contribution of possible large grains on anorthite (blue dashed-dotted line in Fig.~\ref{f:calcium}). The contribution of larger grains does not produce a well
detectable feature.

\subsection{Sulfur}
In the diffuse ISM, sulfur is expected to have a modest depletion (Sect.~\ref{par:elements}). 
We use sulfur in conjunction with iron in the form of troilite, pyrrohtite and pyrite. 
FeS is a likely candidate for a diffuse interstellar environment, due
to its inclusion in GEMS \citep{bradley94, bradley99}. 
The line of sight towards \gx, at distance of $\sim$9\,kpc is likely to cross also molecular 
clouds and this would apply for any source located near the galactic
center. The dust inclusion of sulfur in molecular clouds is still an open issue (Sect.~\ref{par:elements}). Some of the S must be associated to ices and carbon-hydrogen aggregates, while the
rest may be in the form of FeS or atomic gas. Even the sum of all known S-bearing molecules would be unlikely to exceed
few \% of the total S abundance. 
Therefore any significant depletion detected by XRISM or XIFU would naturally point to the role of S in GEMS.  
We note that this amount of S depletion would still not procure visible deviations from the observed total dust spectral energy distribution \citep{kohler14}.

\

\subsection{Titanium and Nickel}
\begin{figure}
\begin{center}
\resizebox{\hsize}{!}{\includegraphics[angle=90]{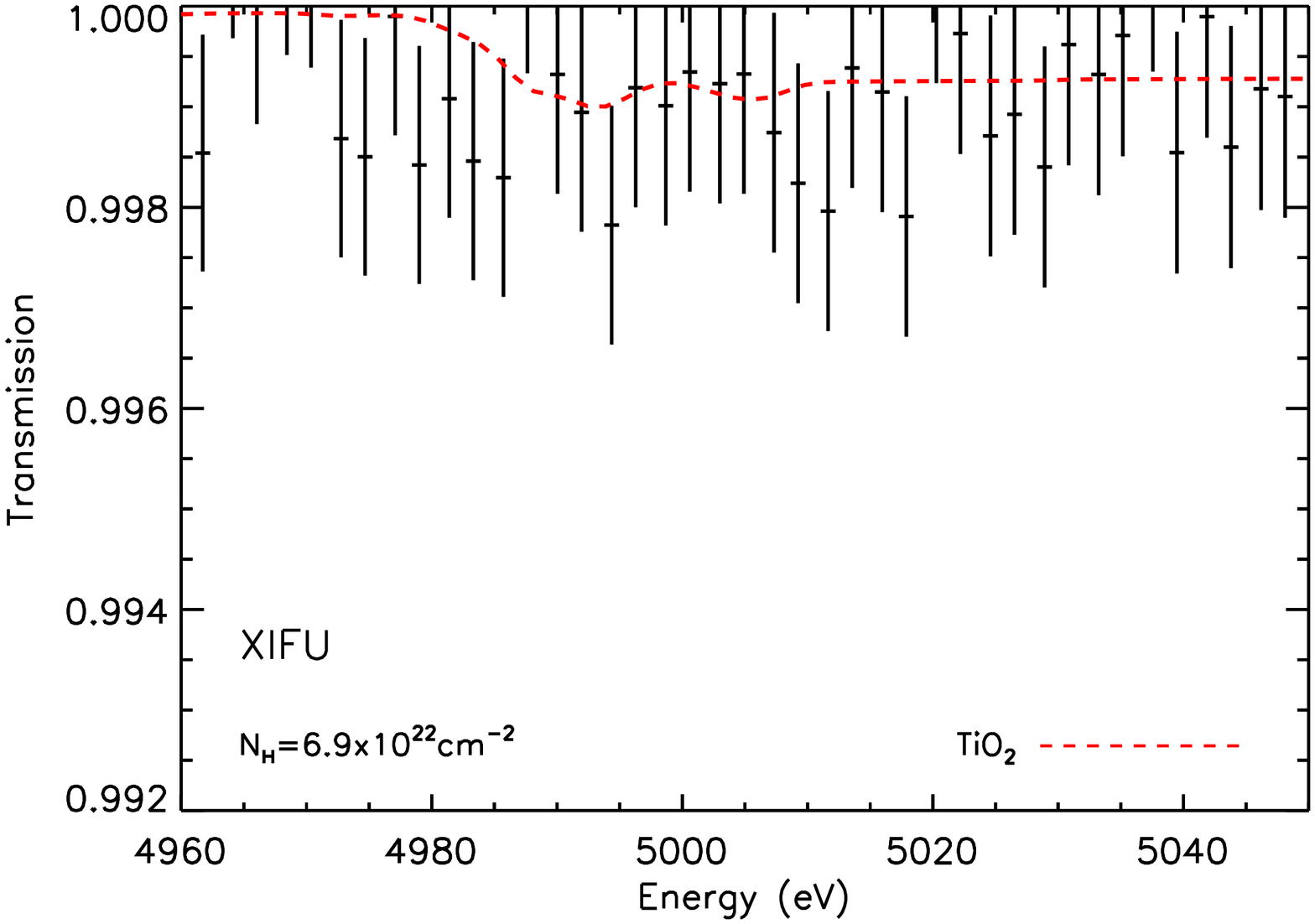}}
\end{center}
\caption{\label{f:titanium} A 500\,ks simulation of the titanium K\,edge, using the XIFU calorimeter, using the bright XRB GX340+00 
($F_{2-10\,{\rm keV}}\sim1.3\times10^{-8}$\,erg\,cm$^{-2}$\,s$^{-1}$) as template. The dust depletion is 100\%. The data have been binned for clarity.} 
\end{figure}

\begin{figure}

\begin{center}
\resizebox{\hsize}{!}{\includegraphics[angle=90]{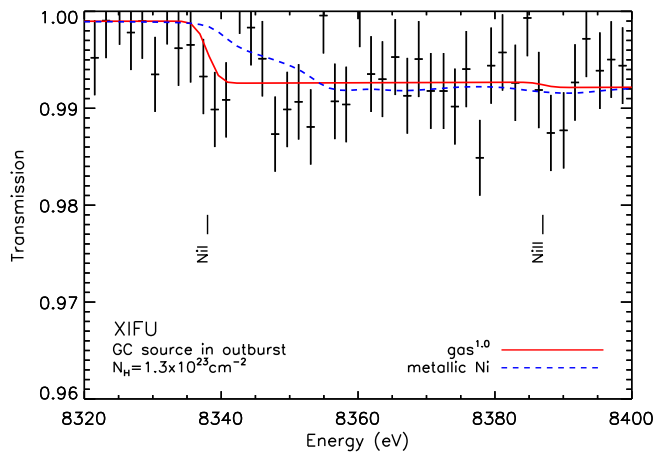}}
\end{center}
\caption{\label{f:nickel} A 300\,ks simulation of the nickel K\,edge, using the XIFU calorimeter, assuming that a highly absorbed source near the GC reaches in outburst the same flux
level as GX340+00 ($F_{2-10\,{\rm keV}}\sim1.3\times10^{-8}$\,erg\,cm$^{-2}$\,s$^{-1}$). The dust depletion is 100\%. The data have been binned for clarity.} 
\end{figure}

Due to its extremely low abundance in the Universe, titanium will be challenging to detect (Fig.~\ref{f:titanium}). 
Nickel is about twenty times more abundant than Ti, however Ni will be also difficult to study (Fig.~\ref{f:nickel}). 
 The large column densities required to produce a Ni edge, will
 cause also strong absorption by iron, whose K absorption edge lies at 7100\,eV, only 1.2\,keV away from the one of nickel. 
 Under the conditions of this simulation, the optical depth of iron will be around 18 times larger than the one of nickel. The net effect is that
 the Ni edge ''sees" a continuum which is much lower than the one of the source, reducing the signal to noise ratio in that feature. 
 Both titanium and nickel are however completely depleted in most ISM environments, therefore even a
 column density estimate will be useful to constrain the abundance of these two elements, which are 
 a product of explosions of both massive stars and white dwarfs.

\section{Conclusion}

In this paper we have shown how improved instrumental sensitivity and resolution will help in understanding new aspects of the composition of ID. Our results can
be summarized as follows:

Future instruments, with characteristics similar to the Arcus mission, will be able to disentangle between the major components of carbon, namely amorphous
carbon (or graphite) and hydrogenated carbon. The effect of minor constituents of C in the ISM (e.g. nano-diamonds and PAH) will be challenging to detect.

Instruments with improved capabilities at higher energies as \at-XIFU or XRISM-{\it Resolve}, will be able to 
study absorption features that, due to their modest opacity, could not be investigated before. Simulations 
show that even a
1-6\% jump in the transmission spectrum will be detected, allowing at least abundances measurements. For 
the low-cosmic
abundance elements investigate in the E>1\,keV band (namely Al, S and Ca), a full characterization (e.g. distinguishing
among various silicate-like compounds) of the dust chemistry
will be likely challenging. However, some main distinctions can be made:
\begin{itemize}
\item It will be possible to distinguish between calcium in carbonates and silicates around the Ca edge. 
\item For both
Ca and Al the dust size distribution of these heavily depleted elements can be determined, with different 
precision depending on the instrument characteristics. 

\item It will be possible  to determine the depletion of sulfur in the ISM. 
This in turn will help to clarify the S inclusion in GEMS, which are
sometimes considered as one of the main forms of silicates in the ISM.     

\end{itemize}

\noindent 
Finally, simulations show that Ti and Ni will be unaccessible to a detailed study even with next 
generation instruments considered
here.

\begin{acknowledgements}
Dust studies at Leiden Observatory are supported also through the Spinoza Premie of the Dutch science agency, NWO. The Netherlands Institute for Space Research is supported
financially by NWO. E.C. and D.R. acknowledge the support of the NWO-VIDI grant 639.042.525. 
We acknowledge SOLEIL for provision of
synchrotron radiation facilities and we would like to thank Delphine Vantelon for
assistance in using the LUCIA beamline and Harald Mutschke for procuring the Al-bearing samples. 
We also thank Alessandra Candian and the anonymous referee for useful comments on the manuscript. 
This research made use of the Chandra
Transmission Grating Catalog and archive (http://tgcat.mit.edu).   
\end{acknowledgements}

\begin{appendix}\vspace{-3cm}
\section{Extinction profiles}\label{par:appendix}
We show here the extinction profiles in transmission, normalized for the continuum, 
of the compounds presented in this paper. Their formula and literature reference is reported in Table~\ref{t:sample}. 
The instruments used for those measurements are
reported in Table~\ref{t:lab_res}.\\ 

\begin{figure}\label{f:a1}
\begin{center}
\resizebox{\hsize}{!}{\includegraphics[angle=90]{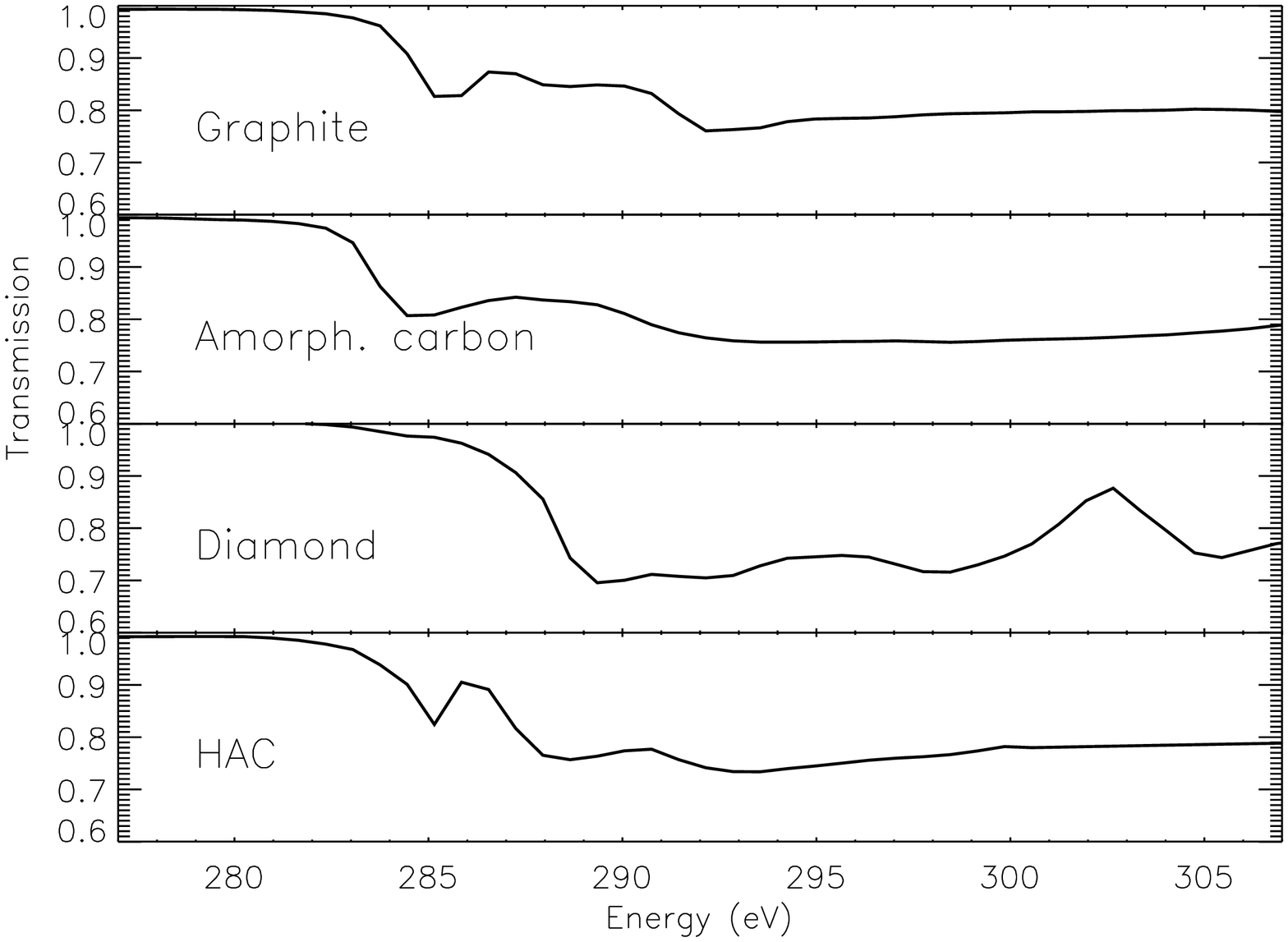}}
\resizebox{\hsize}{!}{\includegraphics[angle=90]{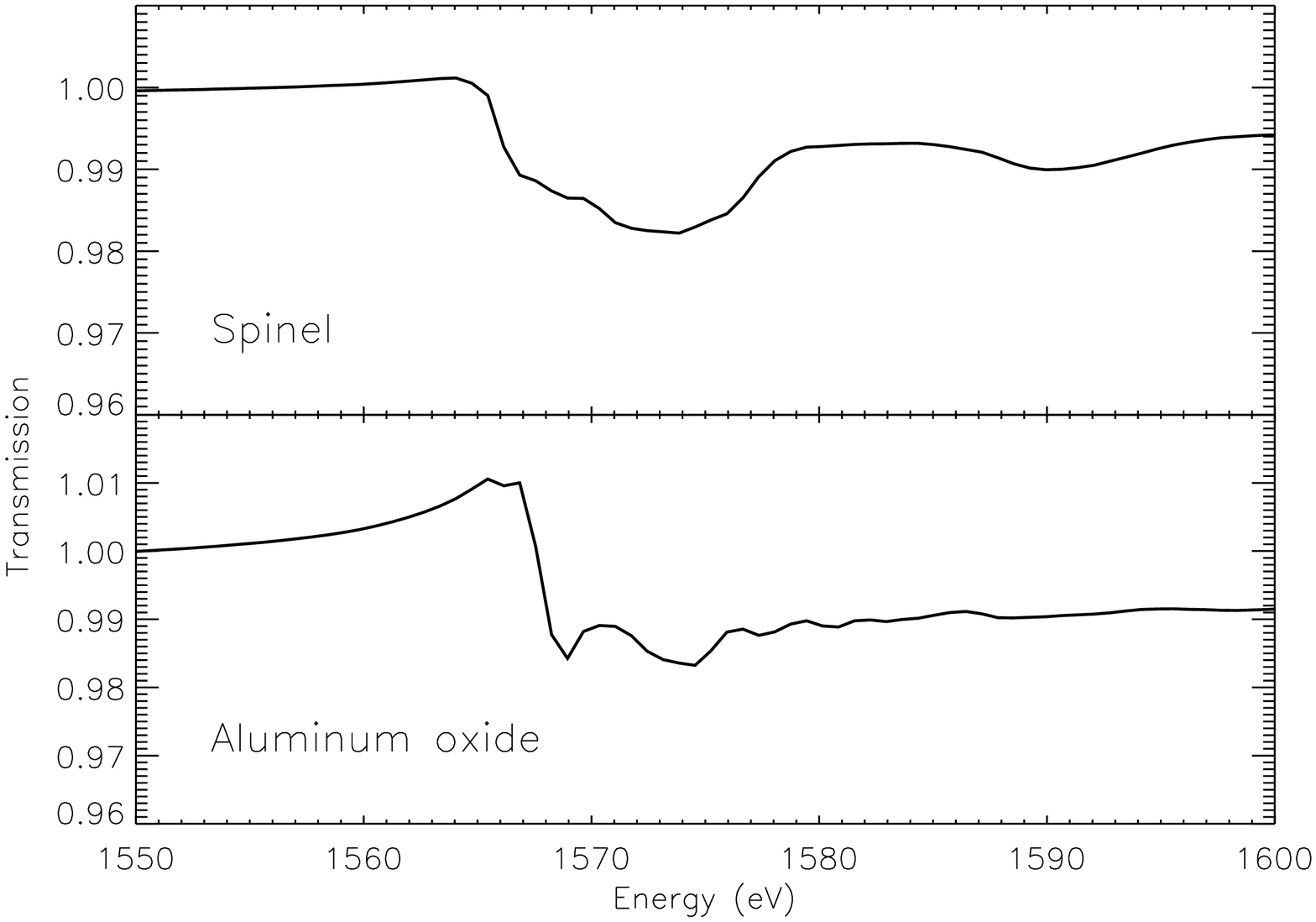}}
\resizebox{\hsize}{!}{\includegraphics[angle=90]{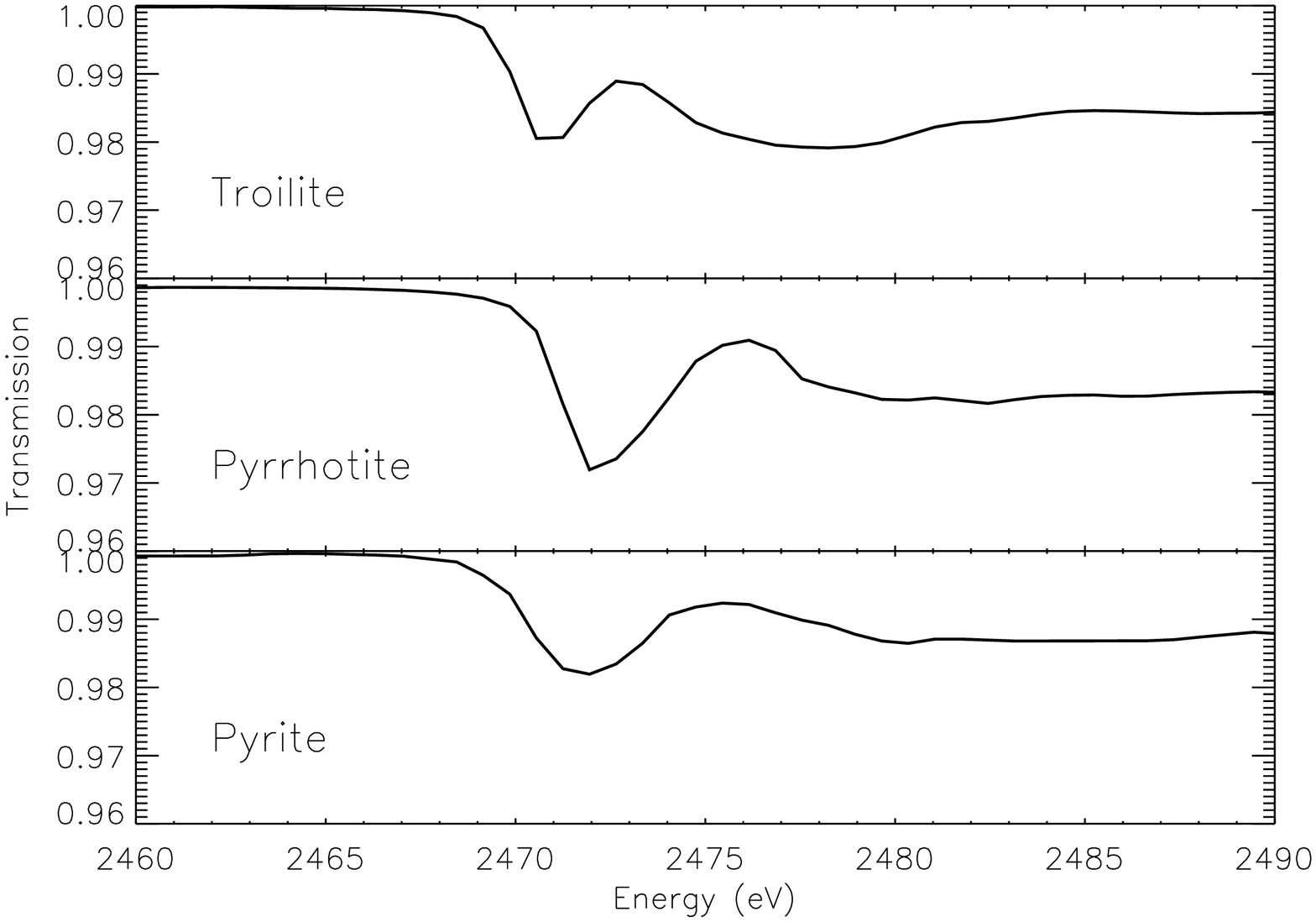}}
\end{center}
\caption{The extinction profiles of the C, Al and S compounds presented in this paper. The dust column densities of the elements are the same used for the simulations. The edge energy is as
reported in the literature as well as the original energy resolution.}

\end{figure}

\begin{figure}\label{f:a2}
\begin{center}
\resizebox{\hsize}{!}{\includegraphics[angle=90]{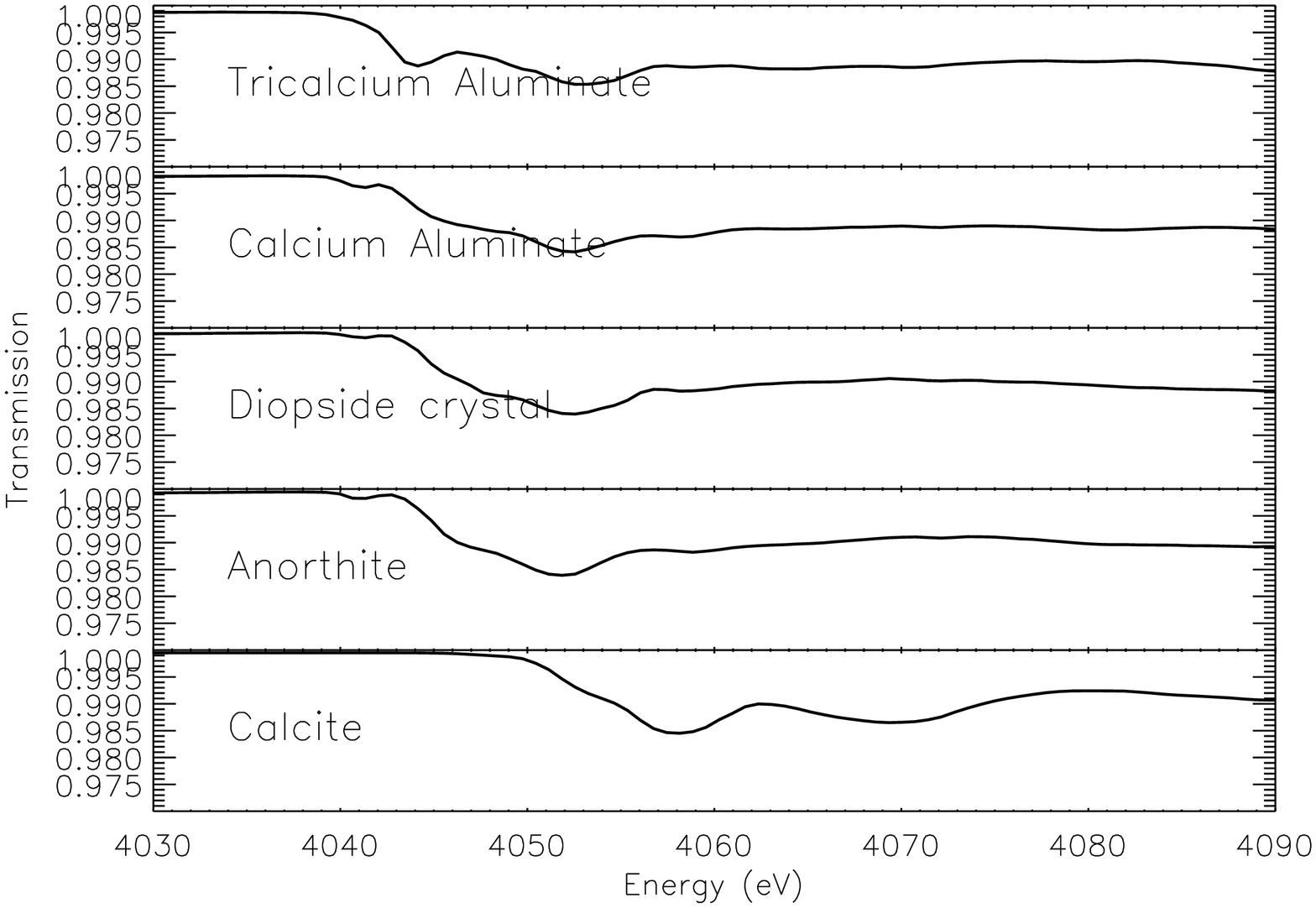}}
\resizebox{\hsize}{!}{\includegraphics[angle=90]{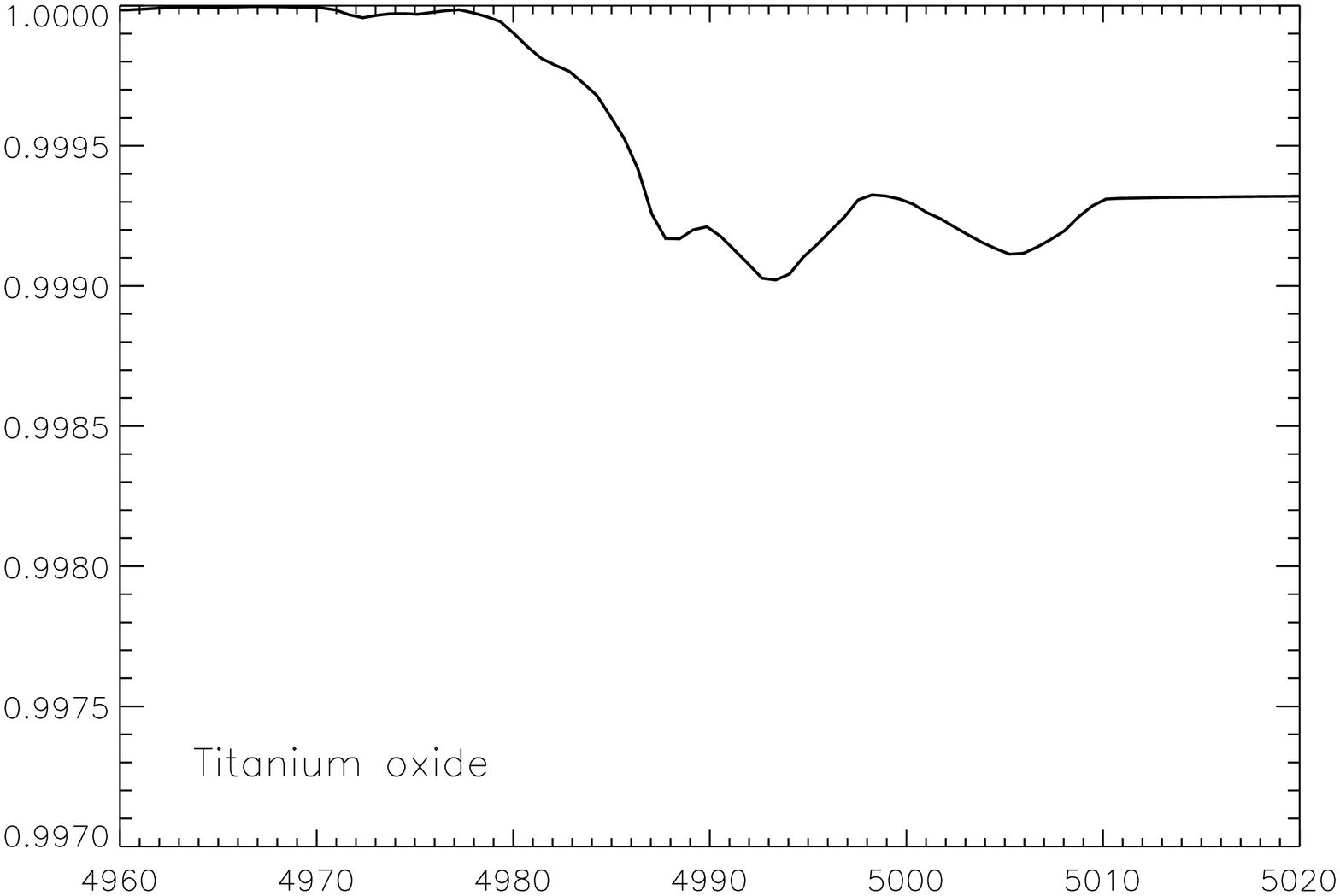}}
\resizebox{\hsize}{!}{\includegraphics[angle=90]{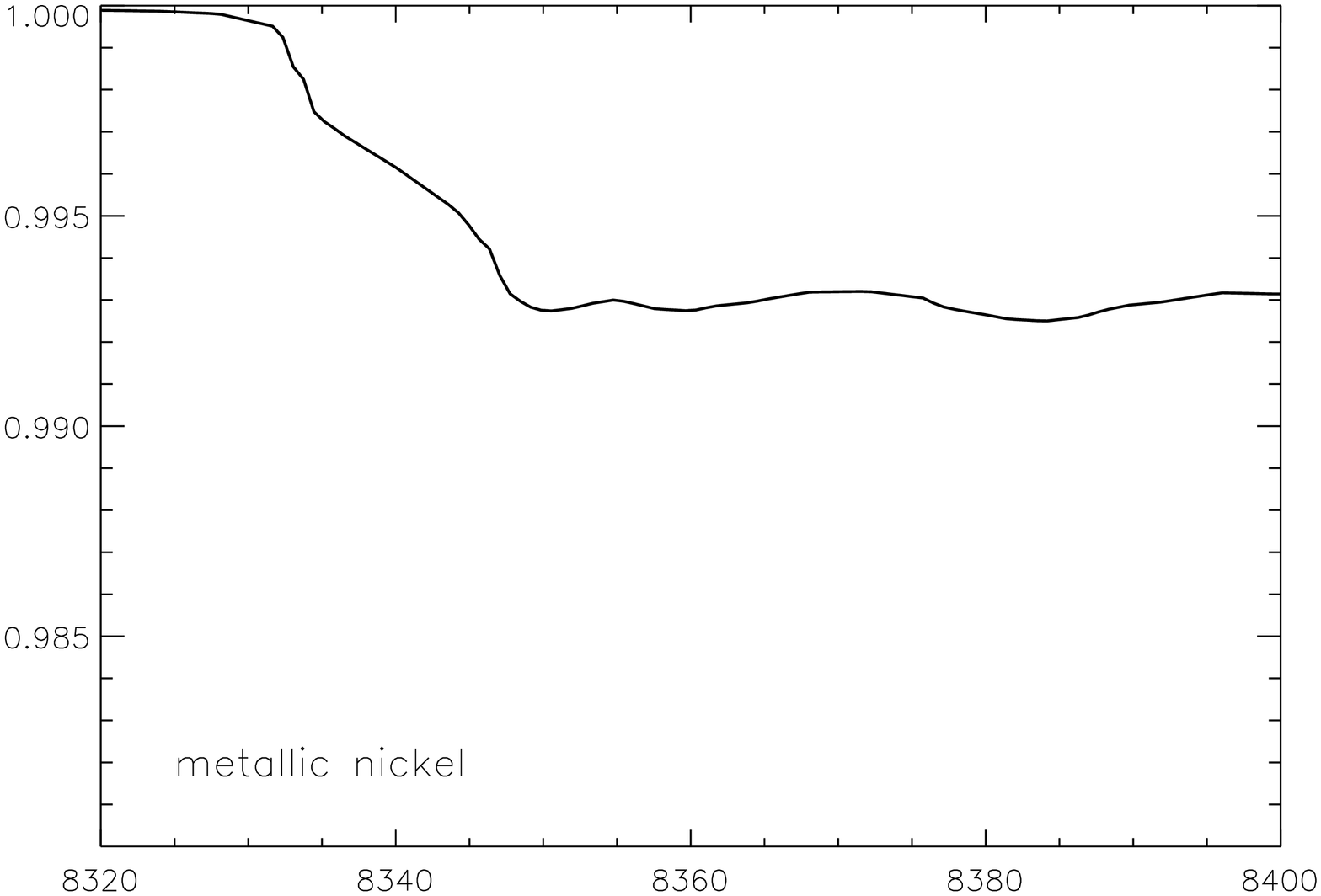}}

\end{center}
\caption{The extinction profiles of the Ca, Ti and Ni compounds presented in this paper. 
The dust column densities of the elements are the same used for the simulations. The edge energy is as
reported in the literature as well as the original energy resolution.}
\end{figure}

\begin{table*}
\caption{\label{t:lab_res} Facilities and resolution of the literature laboratory measurements}
\begin{center}
\begin{tabular}{llll}
\hline
\hline
Ref & {\small facility-beamline} & element & resolution (eV) \\
&&&\\
\citet{gago98} & SSRL-8.2 & C & 0.1 \\
\citet{buijnsters09} & BessyII-SURICAT & C & $\sim$0.05\\
This work & Soleil-LUCIA & Al & 0.25\\
\citet{bonnin02} & ESRF-ID21 & S & 0.3\\
\citet{neuville07} & Soleil-LUCIA & Ca & 0.25\\
\citet{shin13} & PSLII-7D & Ti & $\sim$0.09\\
\citet{vanloon15} & CLS-HXMA & Ni & 0.8\\
\hline
\end{tabular}
\end{center}
\end{table*}

\end{appendix}

%
%-------------------------------------------------------------
%                 A figure as large as the width of the column
%-------------------------------------------------------------
%   \begin{figure}
%   \centering
%   \includegraphics[width=\hsize]{empty.eps}
%      \caption{Vibrational stability equation of state
%               $S_{\mathrm{vib}}(\lg e, \lg \rho)$.
%               $>0$ means vibrational stability.
 %             }
 %        \label{FigVibStab}
 %  \end{figure}
%
%-------------------------------------------------------------
%
\end{document}